\documentclass{article}
\usepackage{psfig}

\def\be{\begin{equation}}
\def\ee{\end{equation}}
\def\bea{\begin{eqnarray}}
\def\eea{\end{eqnarray}}
\begin{document}

\title{Discrete Folding\footnote{Invited talk given by Mark Bowick: to
appear in the Proceedings of the 4th Chia Meeting on ``Condensed
Matter and High Energy Physics.'' (World Scientific, Singapore).}}

\author{ Mark J. Bowick\\Physics Department, Syracuse
University\\Syracuse, NY 13244-1130, USA\\[1em] Philippe Di
Francesco\footnote{Current address: Dept. of Mathematics, Univ. of
North Carolina, Chapel Hill, NC 27599-3250.}, Olivier Golinelli and
Emmanuel Guitter\\CEA, Service de Physique Th\'eorique de Saclay
\\F-91191 Gif sur Yvette Cedex, France} \date{SU-4240-636
\hspace{.25in} Saclay T96/069} \maketitle

\begin{abstract}
Models of folding of a triangular lattice embedded in a discrete space
are studied as simple models of the crumpling transition of
fixed-connectivity membranes. Both the case of planar folding and
three-dimensional folding on a face-centered-cubic lattice are
treated. The 3d-folding problem corresponds to a 96-vertex model
and exhibits a first-order folding transition from a crumpled phase
to a completely flat phase as the bending rigidity increases. 
\end{abstract}
\section{Introduction}

The statistical mechanics of polymers, essentially one-dimensional
objects, has proven to be a rich and fascinating field.\cite{deg,deja}
The success of physical methods applied to polymers arises from
universality --- many of the large-length-scale properties of polymers
are independent of microscopic details such as the chemical identity
of the monomers.  The generalization of polymer statistical mechanics
to membranes, {\em two-dimensional surfaces} fluctuating in some
embedding space, has proven to be even richer and is still under
active development.  In contrast to polymers there are different {\em
universality classes} of membranes distinguished by their long-range
{\em order}.  These classes are the analogues of the well-known {\em
crystalline}, {\em hexatic} and {\em fluid} phases of strictly
two-dimensional systems.\cite{jeru,drnlh} This diversity arises from
the richer geometry of surfaces as compared to lines and the resultant
enlarged space of possible symmetries.

The closest membrane analogue to a polymer is a $2D$ fishnet-like mesh
of monomers (or vertices) with fixed connectivity.  The bonds (links)
are assumed to be unbreakable.  The vertices themselves live in ${\bf
R}^d$, with a physical membrane corresponding to the case $d=3$. Such
a membrane is called a crystalline or polymerized membrane by virtue
of its intrinsic crystalline order. In this talk we will deal only
with {\em phantom} (non-self-avoiding) membranes. In general the
Hamiltonian for a polymerized membrane will have both in-plane elastic
(strain) contributions and out-of-plane bending contributions, since
it can both support shear and fluctuate in the embedding space.  In
Monge gauge the most general effective continuum Hamiltonian at long
wavelength takes the form
\begin{equation}
\label{eqn:contFE}
{\cal H}(h, {\bf u}) = \frac {\kappa}{2} \int d^2\sigma \,
{(\nabla^2h)}^2 + \frac {1}{2} \int d^2\sigma ( 2 \mu \, u_{\alpha
\beta}^2 + \lambda \, u_{\gamma \gamma}^2),
\end{equation}
where $h$ is the height above a reference plane with intrinsic coordinates 
$(\sigma_1,\sigma_2)$, ${\bf u}$ are the phonon modes, 
$u_{\alpha \beta}$ is the strain tensor, $\kappa$ is the bending
rigidity and $\mu$ and $\lambda$ are the bare Lam{\'e} coefficients.
The strain tensor $u_{\alpha \beta}$ measures the difference between
the induced metric and a flat reference metric determined by the
equilibrium configuration of the membrane at rest.
  
The distinctive physics of polymerized membranes arises from an
unavoidable nonlinear coupling between the in-plane strain modes and
the out-of-plane height fluctuations. Integrating out the quadratic
phonon fluctuations by linearizing the strain tensor one finds an
effective long-range interaction between Gaussian curvature
excitations which tends to stiffen the surface on long length
scales.\cite{nepe,jeru} In other words, the effect of undulations on
small length scales is to increase the bending rigidity on longer
length scales. This effect is easily demonstrated --- an ordinary piece
of paper becomes much stiffer with respect to bending deformations
after it is crumpled and then opened up.  More technically, one finds
that the bending rigidity at sufficiently long wavelengths is momentum
dependent $\kappa(q) \sim q^{-\eta}$, with $\eta$ positive. This
scaling behaviour leads to a stable flat phase for polymerized
membranes,\cite{nepe,jeru} with remarkable properties controlled by a
infrared stable fixed point.\cite{arlu,gdlp,agl} For small bending
rigidity (high-temperature), on the other hand, polymerized membranes
are crumpled.\cite{kkn} At some critical bending rigidity (or
equivalently critical temperature) there should, therefore, be a {\em
crumpling} transition from a flat to a crumpled phase.
 
The flat phase of polymerized membranes is characterized by long-range
orientational order in the surface normals. Since long-range order is
unusual in $2D$ systems it is worthwhile to explore as many avenues as
possible to understand its exact nature and origin.  Aronovitz and
Lubensky \cite{arlu} analyzed the flat phase of $D$-dimensional
elastic solids embedded in ${\bf R}^d$, and subject to bending energy,
using an $\epsilon$ expansion about the upper critical (manifold)
dimension $D_{uc}=4$, with fixed codimension $d_c=d-D$. They
determined the renormalization group flows in the space of
dimensionless couplings $\hat \mu = \mu l^{\epsilon}/\kappa^2$ and
$\hat \lambda = \lambda l^{\epsilon}/\kappa^2$, where $l$ is the
renormalization length scale.  They discovered a globally attractive
infrared fixed point describing a flat phase and determined the
scaling dimension $\eta$ and the corresponding exponent $\eta_u$ ($\mu
\sim \lambda \sim q^{\eta_u}$) for the infinite renormalization of the
Lam\'e coefficients.

A revealing extreme limit of polymerized membranes was studied by
David and Guitter.\cite{dagu} This is the limit of infinite
Lam{\'e} coefficients in the Hamiltonian (\ref{eqn:contFE}). Since
the elastic term scales like $q^2$, in momentum space, as compared to
$q^4$ for the bending term, this limit may be regarded as a means of
determining the dominant infrared behaviour of (\ref{eqn:contFE}).  In
this ``stretchless'' limit the strain tensor $u_{\alpha \beta}$ must
vanish and the model is constrained, very much in analogy to a
non-linear sigma model. The $\beta$-function for the suitably rescaled
inverse bending rigidity $\alpha = d/\kappa$ may then be computed in
the large-$d$ limit, yielding
\begin{equation}
\label{eqn:kappabfn}
\beta(\alpha) = q \frac{\partial \alpha}{\partial q} = \frac {2}{d}\ \alpha
		- ( \frac {1}{4 \pi} + \frac {{\rm const}}{d})\ \alpha^2.
\end{equation} 
For $d=\infty$ there is no stable fixed point \cite{pis} and the
membrane is always crumpled.  Eq.~\ref{eqn:kappabfn} reveals, to next
order in $1/d$, an ultraviolet stable fixed point at $\alpha = 8 \pi /
d$, corresponding to a continuous crumpling transition. The critical
exponents associated with the ``flat'' fixed point, discussed above,
may also be determined in the large-$d$ expansion.\cite{gdlp}

\section{Folding}

A natural lattice formulation of the infinite-elastic-constant limit
is the statistical mechanics of folding of a regular
triangular lattice.  A folding in ${\bf R}^d$ of the regular
triangular lattice is a mapping which assigns to each vertex $v$ of
the triangular lattice a position ${\bf X}_v$ in the $d$-dimensional
embedding space ${\bf R}^d$, with the ``metric'' constraint that the
Euclidean distance $|{\bf X}_{v_2}-{\bf X}_{v_1}|$ in ${\bf R}^d$
between nearest neighbours $v_1$ and $v_2$ on the lattice is always
unity.  Under such a mapping, each elementary triangle of the lattice
is mapped onto an equilateral triangle in ${\bf R}^d$.  In general,
two adjacent triangles form some angle in ${\bf R}^d$, i.e. links
serve as hinges between triangles and may be (partially) folded.
\begin{figure}
\centerline{\psfig{figure=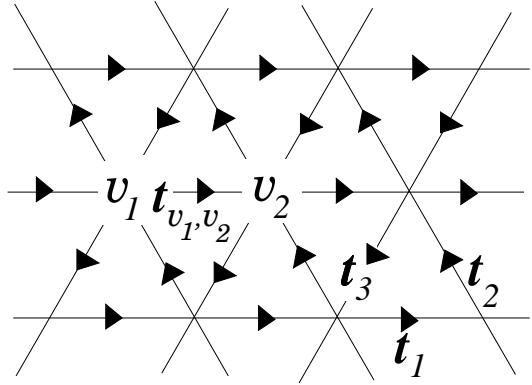,height=2in}}
\caption{The oriented triangular lattice: triangles pointing up
(resp.\ down) are oriented counterclockwise (resp.\ clockwise). The
three tangent vectors ${\bf t}_i$, $i=1,2,3$, have a vanishing sum in
the embedding space.} 
\label{fig:lattice}
\end{figure}
Folding is best described in terms of tangent vectors, which
are link variables defined as follows: we first orient the links of
the lattice as in Fig.\ref{fig:lattice}, with triangles pointing up
(resp.\ down) oriented counterclockwise (resp.\ clockwise), and define
the tangent vector between two neighbours $v_1$ and $v_2$ as the
vector ${\bf t}_{v_1,v_2}= {\bf X}_{v_2}-{\bf X}_{v_1}$ if
the arrow points from $v_1$ to $v_2$. The metric constraint states
that all tangent vectors have {\em unit length}.  With our
choice of orientation, moreover, the three tangent vectors ${\bf t}_i$,
$i=1,2,3$, around each face of the lattice must have {\em vanishing
sum}. This is the basic folding rule: 
\begin{equation}
\label{eqn:foldrule}
{\bf t}_1 +{\bf t}_2+ {\bf t}_3 = {\bf 0}.
\end{equation}
Up to a global translation in ${\bf R}^d$, a folding is therefore a
configuration of unit tangent vectors defined on the links of the
lattice, obeying the folding rule (\ref{eqn:foldrule}) around each
triangle.

Viewed in terms of normals membrane models resemble non-linear sigma
models.  In the lattice version the correspondence is with Heisenberg
spin models. There is one key difference --- apparent in both the
continuum and lattice models. Normal vectors are not arbitrary unit
vectors in the embedding space --- they are {\em constrained} to be
normal to the underlying $2D$ manifold. This means they are actually
Grassmannian sigma models in the continuum \cite{poly} and constrained
Heisenberg models on the lattice. These constraints play a crucial
role in stabilizing an ordered phase. 
   
\section{Planar Folding}

The two-dimensional or {\em planar} ($d=2$) folding problem of the
triangular lattice was first introduced and studied numerically by
Kantor and Jari{\'c}.\cite{kaja} It is clear that the only remnant of
the normal vector degree of freedom is a $Z_2$ spin corresponding to
orientation (up or down). As an Ising spin system the planar folding
model is constrained (as noted above) and is described by an 11-vertex
model. In folding terms these 11 vertices describe the distinct folded
states of a single hexagon of the original triangular lattice. In
terms of the total magnetization $M$ the analogue of the crumpling
transition for this model would be a spin-ordering transition from an
unmagnetized phase ($M=0$) to a magnetized phase ($M \neq 0$).
Analytic progress on planar folding was made
subsequently.\cite{digu1} For planar folding it is easy to check that,
up to a global rotation in the embedding plane, all the link variables
are forced to take their values among a {\em fixed} set of three unit
vectors with vanishing sum. This permits a reformulation of the pure
planar folding problem (with no bending rigidity) as that of the
3--colouring of the links of the triangular lattice: calling the three
fixed vectors blue, white and red, the folding rule
(\ref{eqn:foldrule}) translates into the constraint that the three
colours around each triangle be distinct. This 3--colouring problem
was solved by Baxter \cite{bax} with Bethe Ansatz and transfer matrix
techniques.  His result for the thermodynamic partition function
measures the number of distinct folded configurations $Z_{2d}\propto
q_{2d}^{N_\Delta}$ for a lattice with $N_\Delta$ triangles, in the
limit of large $N_\Delta$.  This gives the folding entropy per
triangle $s_{2d}=\log (q_{2d})$, with \cite{digu1}
\begin{equation}
\label{eqn:exatwod}
{ q_{2d} = \frac {\sqrt{3}}{2 \pi}
\Gamma(1/3)^{3/2} = 1.20872... }
\end{equation}
The $2d$ folding problem has also been studied in the presence of
bending rigidity, which associates an energy to each folded link,
and with a magnetic field coupled to the sum of the normal vectors to the
lattice.\cite{digu2} The model was found to undergo a first order
folding transition.  At zero rigidity and zero magnetic field, the
lattice is in an entropic folded phase ($M=0$). At large enough rigidity
and/or magnetic field, the lattice becomes {\em totally} unfolded
($M= \pm 1$).
\begin{figure}
\centerline{\psfig{figure=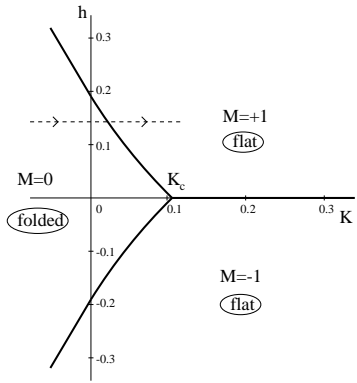,height=2in}}
\caption{The phase diagram for planar folding in the (K,h) plane,
where K is the discrete bending rigidity and h is an external magnetic
field. Three first order lines ${\rm h}={\rm h}_{\rm c}({\rm K}),
-{\rm h}_{\rm c}({\rm K})\; ({\rm K} < {\rm K}_{\rm c})$ and 
${\rm h}= 0 \; ({\rm K} > {\rm K}_{\rm c})$ separate the three phases
${\rm M}= 0, \pm 1$ and meet at the
triple point $({\rm K}_{\rm c},0)$.}
\label{figure:2dphase}
\end{figure}
The phase diagram is shown in Fig.\ \ref{figure:2dphase}.

\section{Three-Dimensional Folding}

\begin{figure}
\centerline{\psfig{figure=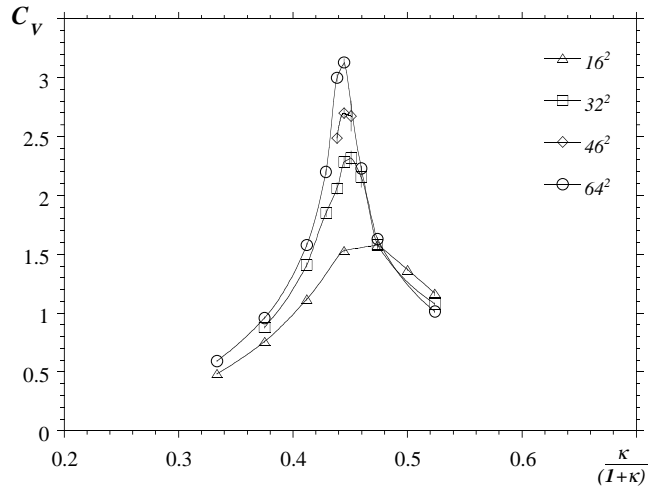,height=2.5in}}
\protect\caption{The specific heat versus bending rigidity from the
Monte Carlo simulation of Ref.~24. The critical exponent
$\alpha=0.4(1)$.}
\label{figure:spec2}
\end{figure}

The existence of a unfolded (magnetized) phase and a first-order
folding transition is fascinating, but one would like to know how
sensitive these features are to the discretization of the space of
local normals. After all, the lower critical dimension for systems
with discrete symmetry is typically less than that for systems with
continuous symmetry. It is conceivable that the transition disappears
altogether if we enlarge the embedding space or that the order of the
transition changes.  In particular there is no rigorous analytic
prediction for the order of the crumpling transition for physical
polymerized membranes. Paczuski, Kardar and Nelson \cite{pkn} found
that critical fluctuations about the mean field solution, in an
$\epsilon=4-D$ expansion, drive the transition first order for
embedding space dimension $d < 219$. While the early numerical
simulations could not rule out a weak first-order
transition,\cite{kane} later more extensive simulations clearly
indicate a continuous transition.\cite{bew,adj,reko,whea,bet,bcfta}
The specific heat plot from Ref.~24 is shown in Fig.\
\ref{figure:spec2}.
To gain some insight into the effect of a finer discretization of the
embedding space we have generalized planar
folding to folding on a $3d$ Face Centred Cubic (FCC)
lattice.\cite{bdgg}

In the general $3$--dimensional folding problem,
the local folding constraint (\ref{eqn:foldrule}) imposes only
that the three tangent vectors around each face be in the same plane
and have relative angles of $2 \pi/3$. This, however, does not
impose any constraint on the {\em relative} positions of the two planes
corresponding to two adjacent faces, which may form some arbitrary 
continuous angle. As opposed to the $2d$ case, this then leads 
to a problem with continuous degrees of freedom.

To define a {\em discrete} model of folding in $3d$, one must allow
only a {\em finite} number of relative angles between adjacent
faces. More specifically, one may also impose that the link variables
themselves take their values among a {\em finite} set of tangent
vectors, now in ${\bf R}^3$.  For symmetry reasons, we took this
set of tangent vectors to be the (oriented) edges of a regular solid
of ${\bf R}^3$, made of equilateral triangles only.
\begin{figure}
\centerline{\psfig{figure=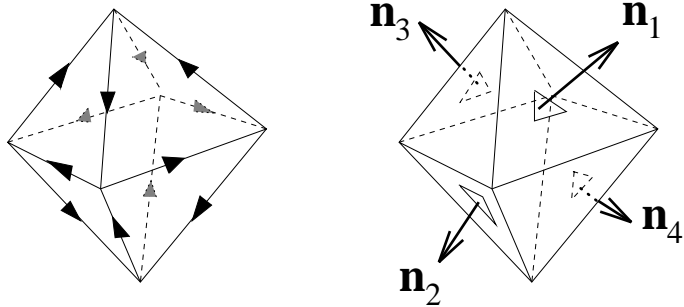,height=2in}}
\caption{The oriented octahedron: the edges around each face form
triplets of tangent vectors with vanishing sum. The four normal
vectors ${\bf n}_i$, $i=1,2,3,4$, are represented on the corresponding
outward oriented faces.}
\label{figure:octaor}
\end{figure}
There are only three regular solids in ${\bf R}^3$ made of equilateral
triangles: the tetrahedron, the octahedron and the icosahedron.  The
edges of the tetrahedron (resp.\ icosahedron), however, cannot be
consistently oriented such that the corresponding tangent vectors
satisfy (\ref{eqn:foldrule}) around each face.  This is because each
vertex is surrounded by an odd number 3 (resp.\ 5) of triangles.  There
is no such problem for the octahedron, as shown in
Fig.~\ref{figure:octaor}.  The $12$ links of the octahedron are
oriented consistently to form $8$ triplets of tangent vectors with
vanishing sum, corresponding to the $8$ faces of the octahedron.
One may therefore consider the restricted $3d$
``octahedral folding'' problem, where the tangent vectors are chosen
from the set of 12 edge vectors of a regular oriented
octahedron. In the folding process, the folding rule (\ref{eqn:foldrule})
imposes that the three links of a given face on the original
triangular lattice are mapped onto one of the $8$ triplets of tangent
vectors above. For a given triplet, the triangle can still be in $3!$
states corresponding to the $3!$ permutations of the three edges.
Each triangle can therefore be in one of $48=8\times 6$ states.

The $8$ faces of the octahedron can be labelled as follows: we consider
for each face its normal vector, pointing outward or inward
according to the orientation of its tangent vectors on the octahedron
(see Fig.~\ref{figure:octaor}). On the octahedron there are four 
alternating outward and inward oriented faces. The
normal vectors to opposite faces are equal. This defines a set of
four vectors ${\bf n}_1$, ${\bf n}_2$, ${\bf n}_3$ and ${\bf
n}_4$, which furthermore satisfy the sum rule ${\bf n}_1+{\bf n}_2+{\bf
n}_3+{\bf n}_4={\bf 0}$.  Each face is labelled by its orientation
(outward or inward) and its normal vector (1, 2, 3 or 4).  
\begin{figure}
\centerline{\psfig{figure=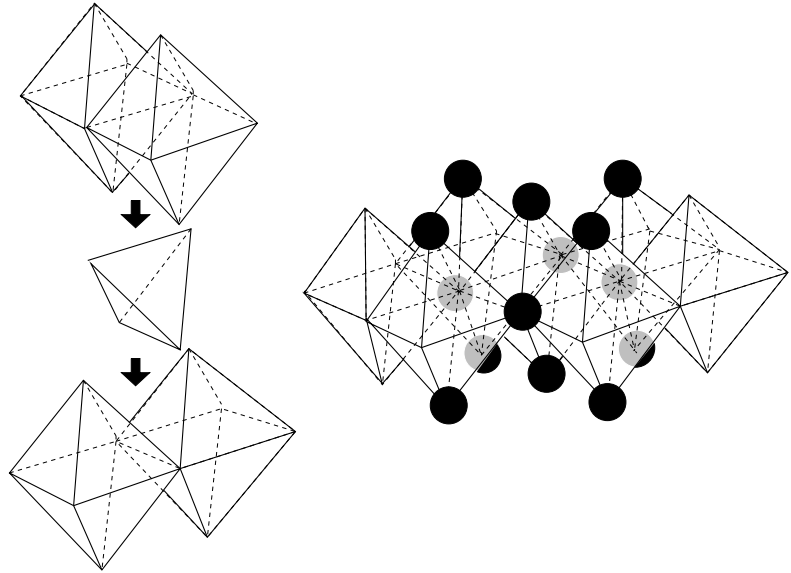,height=3in}}
\caption{The FCC lattice viewed as a packing of $3d$ space
with octahedra and tetrahedra.}
\label{figure:fcc}
\end{figure}

The 12 oriented edge vectors of the octahedron are actually identical
to the 6 edge vectors of a tetrahedron, now taken with both
possible orientations.  The four normal vectors above are also the
normals to this tetrahedron.  For each folding map, the image of the
folded lattice in ${\bf R}^3$ lies therefore on a 3d FCC lattice,
which consists of a packing of space by octahedra complemented by
tetrahedra, as shown in Fig.~\ref{figure:fcc}.  In this respect, the
``octahedral folding'' problem simply corresponds to discretizing the
embedding space as an FCC lattice.

\subsection{96--vertex Model}

When stated in terms of tangent vectors, the $3d$ ``octahedral
folding'' problem involves three types of constraints: face, link and
vertex constraints.  The first constraint, around each {\it face},
imposes that the three tangent vectors of a given triangle form one of
the $8\times 6$ (ordered) triplets with vanishing sum. The second
constraint, on each {\it link}, arises because two adjacent triangles
share a common tangent vector. Given the state of one triangle, any
adjacent triangle has one of its tangent vectors already fixed and
thus is left with only $4=48/12$ possible states.  
\begin{figure}
\centerline{\psfig{figure=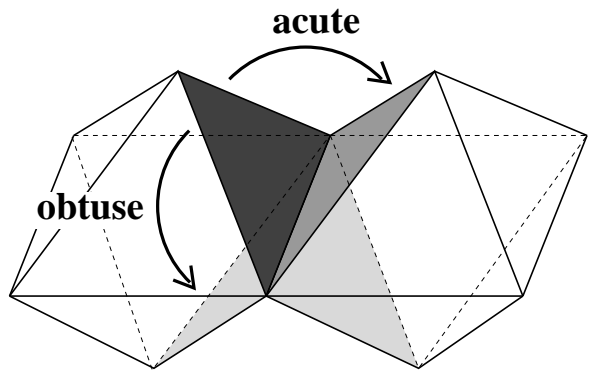,height=2in}}
\caption{The four possible folding angles between two adjacent
triangles. The neighbour of the dark triangle may (i) be itself on top
of the dark triangle (complete fold), (ii) occupy the symmetric
position in the same plane (no fold), (iii) lie on the same octahedron
(i.e. form an obtuse angle) or (iv) lie on the same tetrahedron
(i.e. form an acute angle).}
\label{figure:acutob}
\end{figure}
They correspond simply to the four values for the relative angle
between two neighbouring triangles, i.e. the angle between the normal
vectors, depicted in Fig.~\ref{figure:acutob}.  These four values are
0 (no fold: the triangles are side by side), $180^\circ$ (complete fold:
the triangles are on top of each other), $\cos^{-1}(-1/3)\sim 109^\circ28'$
(fold with acute angle: the two triangles lie on the same tetrahedron)
and $\cos^{-1}(1/3)\sim 70^\circ32'$ (fold with obtuse angle: the triangles
lie on the same octahedron).  Finally, there is a third constraint on
the six successive folds around each {\em vertex} of the lattice:
after making one loop, the same tangent vector must be recovered.
Since the ``metric constraint'' is local, there are actually no
constraints other than these three (face, link and vertex)
constraints.
  
In the study of the $2d$ folding (3--colouring) problem, the face and
link constraints are taken into account by going to ${\bf Z}_2$ spin
variables $\sigma_i$ defined on the faces of the lattice. Ordering the
colours cyclically, the spin is $+1$ (resp.\  $-1$) if the colour
increases (resp.\ decreases) from one link to the neighbouring one on
the triangle, oriented counterclockwise.  In this language, the actual
folds take place exactly on the domain walls of the spin
variable. Instead of having a ${\bf Z}_3$ colour variable per link,
one is left with a ${\bf Z}_2$ spin variable per triangle. The vertex
constraint translates into a constraint on the six spins
$\sigma_1,...,\sigma_6$ around each vertex of the lattice, namely that
$\sum_{i=1}^6\sigma_i=0$ mod 3. This leads to 22 possible local spin
configurations around each vertex, or equivalently, after removing the
global ${\bf Z}_2$ degeneracy of reversal of all spins, to an
11--vertex model on the lattice.\cite{digu1}

One may proceed in the same way for the $3d$ ``octahedral folding''
and account for the face and link constraints by expressing folded
configurations in terms of two ${\bf Z}_2$ variables on the triangles.
These variables will indicate the relative states of successive links
around the face. One may then count the number of allowed hexagonal
configurations around a vertex: this leads to a 96--vertex
model. These vertices and the corresponding rules on the ${\bf Z}_2$
variables will be given below.

Let us label the $12$ edges of the octahedron as follows: each edge is
shared by two adjacent faces, one outward and one inward oriented (see
Fig.\ \ref{figure:octaor}). We label the edges by the indices $(i,j)$,
$1 \leq i \neq j \leq 4$, when the normal vector for the outward face
is ${\bf n}_i$ and the one for the inward face is ${\bf n}_j$.   
There are $12$ such couples
$(i,j)$.  
\begin{figure}
\centerline{\psfig{figure=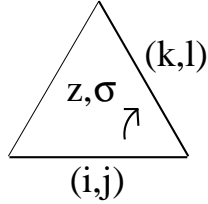,height=1in}}
\caption{The transition from a link $(i,j)$ to a subsequent link
$(k,l)$ is described by the two ${\bf Z}_2$ face variables $z$ and
$\sigma$.}
\label{figure:onestep}
\end{figure}
Consider now an elementary triangle of the lattice.  Starting from one
of its links $(i,j)$ the subsequent link $(k,l)$ counterclockwise must
share a face with $(i,j)$ on the octahedron.  This leads to the $4$
following possibilities, labelled by the ${\bf Z}_2$ face variables
$z, \ \sigma \in \{\pm 1\}$: 
\begin{equation}
\label{eqn:fourchoice} 
\begin{array}{cl}
 z=+1: (i,j) \rightarrow (i,l), \ l \neq j, & 
\epsilon_{ijl} = -\sigma= \pm 1\\
 z=-1: (i,j) \rightarrow (k,j), \ k\neq i,  & 
\epsilon_{ijk} = +\sigma= \pm 1,\\
\end{array}
\end{equation}
where $\epsilon_{ijk}=\sum_l \epsilon_{ijkl}$ is defined in
terms of the totally antisymmetric tensor $\epsilon_{ijkl}$, equal to
the signature of the permutation $(ijkl)$ of $(1234)$.  The value
$z=+1$ (resp.\ $z=-1$) indicates that the two tangent vectors share an
outward oriented (resp.\ inward oriented) face on the octahedron.  The
spin variable $\sigma$ takes the value $+1$ (resp.\ $-1$) if $(k,l)$
follows (resp.\ precedes) $(i,j)$ on their common (oriented) face of
the octahedron.  The variable $\sigma$
also indicates whether the normal vector to the triangle (in the
embedding space ${\bf R}^3$) is parallel ($\sigma=+1$) or antiparallel
($\sigma=-1$) to the corresponding normal vector of the octahedron.

Considering now two neighbouring triangles, the $4$ possible relative
values $z_2/z_1$ and $\sigma_2/\sigma_1$ indicate which type of fold
they form.  The domain walls for the $z$ variable are the location of
the folds which are either acute or obtuse, whereas those for the
$\sigma$ variable are the location of the folds which are either
complete or obtuse.  The superposition of these two types of domain
walls fixes the folding state of all the links, specifying the folding
state of the lattice up to a global orientation.

The use of $z$ and $\sigma$ variables instead of the $12$ $(i,j)$
variables incorporates the face and link constraints. As in the $2d$
case, the vertex constraint is more subtle.  Nevertheless, one can
count the number of possible configurations around a vertex satisfying
this constraint, i.e. the number of possible folded states of an
elementary hexagon.  Indeed the mapping (\ref{eqn:fourchoice}) may be
represented by a $12\times 12$ connectivity matrix $M_{(i,j),(k,l)}$
with $i\ne j$ and $k\ne l$: 
\begin{equation}
\label{eqn:connectivity}
{M_{(i,j),(k,l)}= \delta_{ik}+\delta_{jl}-2\, \delta_{ik}\delta_{jl} \
.}
\end{equation}
This matrix acts as a transfer matrix between two successive
internal links of the hexagon.  The number of configurations of a
hexagon is simply given by: ${\rm Tr} (M^6)=4608$, where the
trace guarantees that the same link variable is recovered after one
loop.  These $4608$ configurations count as distinct all the foldings
which are related by a global change of orientation of the hexagon in
embedding space.  The order of the resultant degeneracy is $48$,
corresponding to $12$ choices for the first tangent on the octahedron
times $4$ for the choice of the second from among its 4 neighbours
(this latter choice corresponds to the $4$ choices of the $z$ and
$\sigma$ variables on the corresponding triangle).  This leaves us
with $4608/48 = {\bf 96}$ distinct configurations.

\subsection{Vertex Rules}
\begin{figure}
\centerline{\psfig{figure=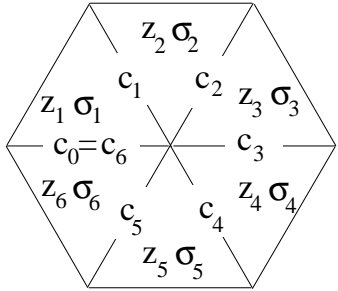,height=1.5in}}
\caption{The six $z_i$ and $\sigma_i$ variables around a given vertex,
and the colours $c_i$ of the interior links.}
\label{figure:spifig}
\end{figure}
Consider an elementary hexagon of the triangular lattice.  As shown in
Fig.~\ref{figure:spifig} each of its 6 triangles is labelled by $z$
and $\sigma$ variables. Let us also assign a colour $c_i$ to each
link: $c_i=c_0 + \sum_{j=1}^{i} \sigma_j \ {\rm mod} \ 3$.
Geometrically each colour corresponds to one of the three orthogonal
planes of the target octahedron (comprising four links).  The vertex
constraint arises from the requirement that six applications of the
link mapping of Eq.(\ref{eqn:fourchoice}) be the identity.  By
relating the link mappings to the elements of the tetrahedral group
$A_4$ one finds two folding rules.\cite{bdgg} The {\em first folding
rule}:
\begin{equation}
\label{eqn:ffr}
{\sum_{i=1}^6 \sigma_i = 0\ {\rm mod}\ 3}
\end{equation}
for the six spins around the central vertex of the hexagon is
identical to the planar folding rule.  
In contrast with the $2d$ situation, the restriction (\ref{eqn:ffr})
is not the only constraint.  There is a {\em second folding rule}
\begin{equation}
\label{eqn:sfr}
{\sum_{i=1}^{6}\frac{1}{2}(1 - z_iz_{i+1})\delta(c_i,c \ {\rm mod} \ 3)
 = 0 \ {\rm mod} \ 2\, : \quad c=1,2 \ .}
\end{equation}
The two folding rules (\ref{eqn:ffr}) and (\ref{eqn:sfr}) characterise
the vertex constraint entirely.
\begin{figure}
\centerline{\psfig{figure=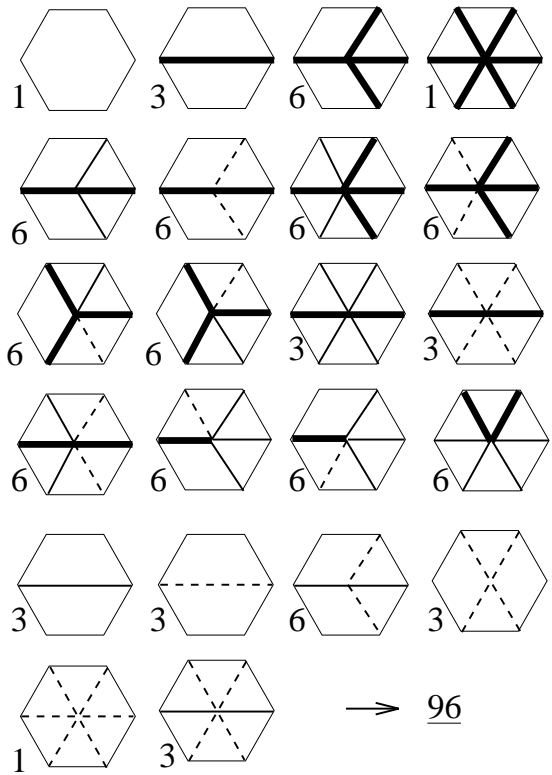,height=4in}}
\caption{The 96 vertices satisfying the two folding rules
\ref{eqn:ffr} and \ref{eqn:sfr}:
no line corresponds to no fold, a thick line corresponds to a complete fold,
a thin line  corresponds to a fold with obtuse angle and a 
dashed line corresponds to a fold with
acute angle. The degeneracy of each vertex under cyclic
permutations of the links is indicated.}
\label{figure:ninetysix}
\end{figure}
One finds $384=96 \times 4$ vertex configurations (there is a
$4$--fold global degeneracy under reversal of $z$ or $\sigma$).  The
$96$ folding vertices are displayed in Fig.~\ref{figure:ninetysix}
with the following conventions: no line corresponds to no fold; a
thick line corresponds to a complete folding ($180^{\circ}$, flip of
$\sigma$ only); a thin line corresponds to a fold with obtuse angle
($\cos^{-1}(1/3)\sim 70^\circ 32'$ between normal vectors, flip of
both $\sigma$ and $z$) and finally a dashed line corresponds to a fold
with acute angle ($\cos^{-1}(-1/3)\sim 109^\circ 28'$, flip of $z$
only).  The degeneracy of each vertex under cyclic permutations of the
links is also indicated.
\begin{figure}
\centerline{\psfig{figure=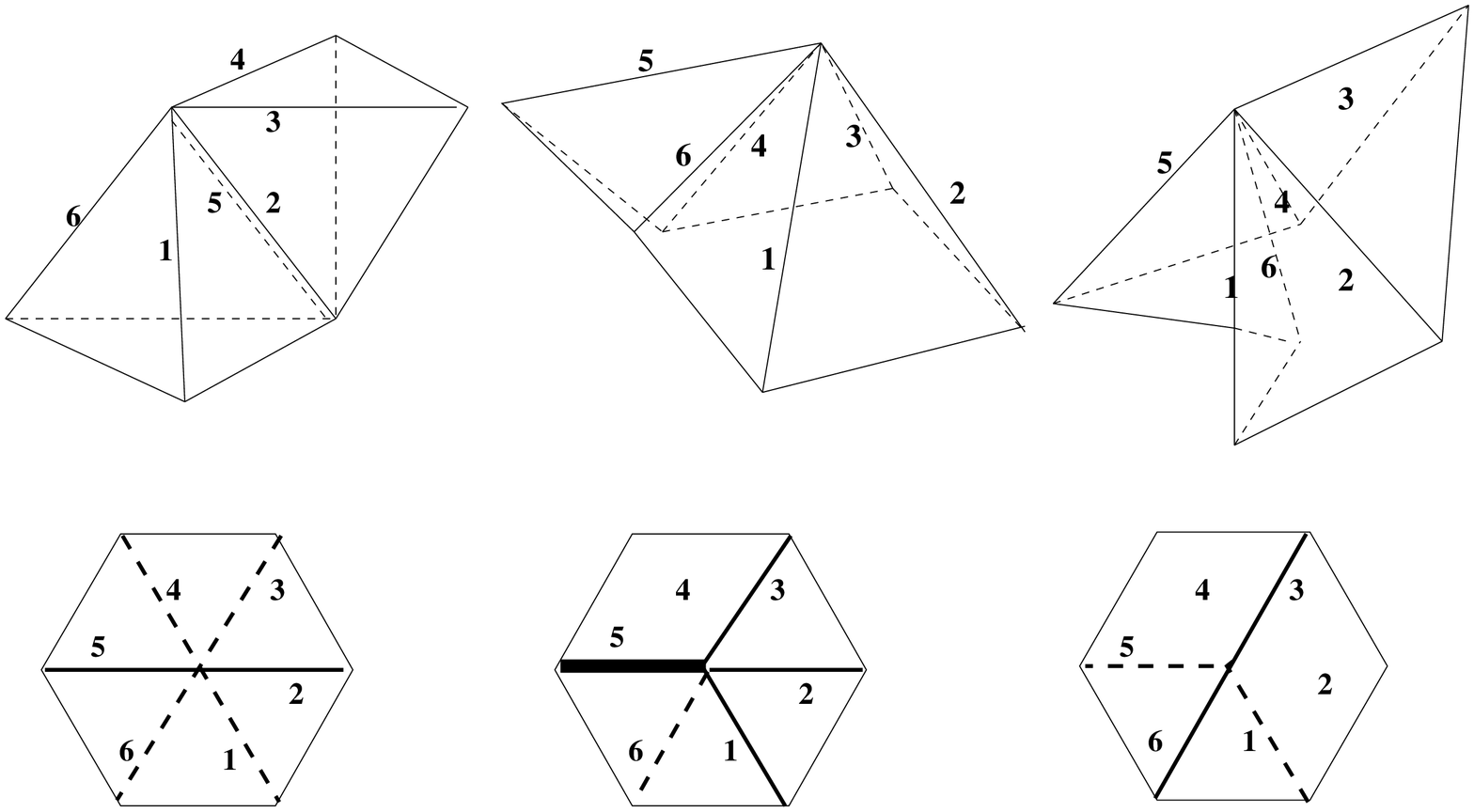,height=2in}}
\caption{Examples of $3d$ octahedral foldings of
an elementary hexagon and the corresponding vertices of
Fig.~\ref{figure:ninetysix}.}
\label{figure:exvert}
\end{figure}
In Fig.~\ref{figure:exvert} we display a few examples of
vertices and the corresponding foldings in $3d$ space.
\begin{figure}
\centerline{\psfig{figure=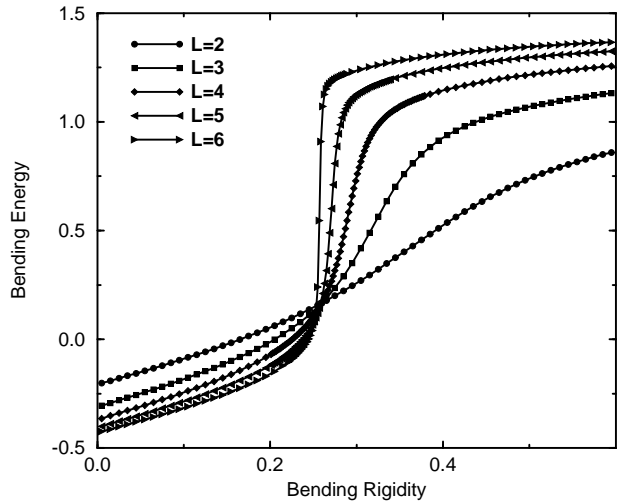,height=3in}}
\caption{Bending Energy vs rigidity for $3d$ folding from a 
transfer matrix calculation on strips of width $L=2,3,4,5$ and $6$.}
\label{figure:3denergy}
\end{figure}
The $3d$ octahedral folding problem is thus precisely defined as a
$96$--vertex model with the vertices of Fig.~\ref{figure:ninetysix}.
Although the entropy of the $3d$ problem cannot be evaluated exactly
it can be determined by a numerical transfer matrix
calculation to be $q_{3d} \approx 1.43(1)$.\cite{bdgg} One can also
derive various exact bounds on the entropy for $3d$ folding by
relating it to dressed 3--colouring and $2d$ folding in an external
staggered magnetic field.\cite{bdgg} The best bounds derived in this
way, expressed numerically to three decimal places, are $1.436 \leq
q_{3d} \leq 1.589$.

We have also examined the phase diagram of $3d$ folding with
bending energy. As for planar folding one finds a first-order phase
transition to an ordered state at a critical value of the bending
rigidity. The bending energy versus rigidity is shown in 
Fig.~\ref{figure:3denergy} for infinite strips of width $L=2,3,4,5$ and
$6$. One sees quite clearly the emergence of a non-zero latent
heat as the system size increases. This is confirmed by a finite-size
scaling analysis of the peak of the specific heat plot: it grows linearly with
system size, as characteristic of a first-order transition.

The planar and $3d$ folding models described here have also been
studied in the hexagon approximation of the cluster variation
method.\cite{cgp} For planar folding this allows the incorporation of
defects in the lattice and the study of the crossover from the pure
Ising model to the pure folding problem.  For $3d$ folding these
authors also add a symmetry-breaking field to the model and find a
first-order transition to a flat phase for any value of the symmetry
breaking field.

In conclusion folding is a rich problem providing considerable insight
into the phase structure of fluctuating membranes.

\section*{Acknowledgments}
The research of M.B. was supported by the Department of Energy, USA,
under contract No. DE-FG02-85ER40237. M.B. would also like to thank
the Service de Physique Th\'eorique of Saclay for its kind
hospitality during a visit in which some of the work described here
was done as well as Alberto Devoto and the organizers of the wonderful
meeting at Chia (September 4-8, 1995).

\end{document}